\documentclass[amsmath,amssymb,amsfonts,aps,prl,superscriptaddress,bibnotes,showpacs,showkeys,longbibliography,notitlepage,10pt,twocolumn]{revtex4-2}
\pdfoutput=1

\usepackage{mathtools}
\usepackage{epsfig}
\usepackage[english]{babel}
\usepackage{color}
\usepackage{subcaption}
\usepackage{hyperref}
\usepackage{lipsum}
\usepackage{xcolor}
\usepackage{amsthm}
\usepackage{multirow}
\usepackage[utf8]{inputenc}

\AtBeginDocument{%
    \newwrite\bibnotes
    \def\bibnotesext{Notes.bib}
    \immediate\openout\bibnotes=\jobname\bibnotesext
    \immediate\write\bibnotes{@CONTROL{REVTEX41Control}}
    \immediate\write\bibnotes{@CONTROL{%
    apsrev41Control,author="08",editor="1",pages="1",title="0",year="1"}}
     \if@filesw
     \immediate\write\@auxout{\string\citation{apsrev41Control}}%
    \fi
}%

\begin{document}
\title{The Energy-Momentum Tensor of Electromagnetic Waves in Media and the Solution to the Abraham-Minkowski Dilemma}

\author{Gui-Xiong Liang}
\affiliation{Cangwu Wangfu Education committee office, Wuzhou city, China, District 543004}

\date{\today}

\begin{abstract}
We find that the energy-dynamic tensor of electromagnetic wave in medium defined by Lagrange density is very similar to that of ordinary fluid, and density and pressure can be defined similarly. The pressure here is different from the traditional light pressure. On this basis, we discuss the average momentum, pressure and Bernoulli effect of electromagnetic wave. After considering the contribution of the interaction term, we obtain a new conserved energy-momentum tensor, from which Minkowski momentum and Abraham momentum are derived simultaneously. The difference between them is whether the contribution of pressure is included in momentum. When the pressure significantly affects the experimental results, the experiment will support Minkowski momentum. On the contrary, Abraham momentum is supported. The findings of this paper provide a new solution to the Abraham-Minkowski Dilemma.
\end{abstract}

\maketitle
Two completely different forms of the energy-momentum tensor of electromagnetic waves in media were first proposed by Minkowski\textcolor[rgb]{0.184313725,0.188235294117647,0.564705882}{\cite{Minkowski1,Minkowski2}} and Abraham\textcolor[rgb]{0.184313725,0.188235294117647,0.564705882}{\cite{Abraham1,Abraham2}}, but later people put forward more different views\textcolor[rgb]{0.184313725,0.188235294117647,0.564705882}{\cite{LinZonghan,Groot,Peierls,Stephen}}. According to Minkowski tensor, the momentum of electromagnetic wave in the medium with refractive index $n$ will be $n$ times that in free space; But according to Abraham tensor, it is $1/n$. These two completely different results caused the Abraham-Minkowski Dilemma. For more than a century, both sides of the debate have conducted in-depth theoretical discussion\textcolor[rgb]{0.184313725,0.188235294117647,0.564705882}{\cite{2007New,Stephen,2020Dielectric}} and experimental verification\textcolor[rgb]{0.184313725,0.188235294117647,0.564705882}{\cite{JONES,1973Radiation,A1980Walker,Campbell2005,Weilong2008}}. Brevik\textcolor[rgb]{0.184313725,0.188235294117647,0.564705882}{\cite{Brevik1979}} and Pfeifer\textcolor[rgb]{0.184313725,0.188235294117647,0.564705882}{\cite{Pfeifer}} reviewed the controversy in 1979 and 2007 respectively. Pfeifer believes that the dispute has been resolved, and Leonhardt\textcolor[rgb]{0.184313725,0.188235294117647,0.564705882}{\cite{Leonhardt}} believes that this problem is still the research object of theory and experiment. \par

In the face of the above debate for more than a century, we believe that referring to the general theory of relativity\textcolor[rgb]{0.184313725,0.188235294117647,0.564705882}{\cite{Dirac,Wheeler}}, it is appropriate to define the total energy-momentum tensor of the electromagnetic wave in the medium by:
\begin{equation}
    \label{eq:1}
 T^{\mu  \nu }=\frac{-2}{\sqrt{-g}}\frac{\delta (\mathcal{L} \sqrt{-g})}{\delta  g_{\mu  \nu }}
\end{equation}
In the presence of a medium, the Lagrange density can be expressed as $\mathcal{L}=\, _F\mathcal{L}+\, _I\mathcal{L}+\, _m\mathcal{L}$, where $\, _F\mathcal{L}=-\frac{1}{4\mu _0}F^{\mu  \nu }F^{\alpha  \beta }g_{\mu  \alpha }g_{\nu  \beta }$ is the Lagrange density of the electromagnetic field\textcolor[rgb]{0.184313725,0.188235294117647,0.564705882}{\cite{LiangCanbin2006}}, $\, _I\mathcal{L}=A^{\mu }j^{\nu }g_{\mu  \nu }$ is the interaction term between the electromagnetic field and the medium, and $\, _M\mathcal{L}$ is the
Lagrange density determined by the nature of the medium itself. The total energy-momentum tensor $T^{\mu  \nu }$ is naturally divided into three parts, where $\, _FT^{\mu  \nu }=\frac{-2}{\sqrt{-g}}\frac{\delta (\, _F\mathcal{L} \sqrt{-g})}{\delta  g_{\mu  \nu }}$ belongs to pure electromagnetic field, $\, _IT^{\mu  \nu }=\frac{-2}{\sqrt{-g}}\frac{\delta (\, _I\mathcal{L} \sqrt{-g})}{\delta  g_{\mu  \nu }}$ is the contribution of the interaction term,
and $\, _mT^{\mu  \nu }=\frac{-2}{\sqrt{-g}}\frac{\delta (\, _m\mathcal{L} \sqrt{-g})}{\delta  g_{\mu  \nu }}$ belongs to the medium itself. From Eq.\textbf{(}\ref{eq:1}\textbf{)} and $\, _F\mathcal{L}$, we get the energy-momentum tensor of pure electromagnetic field is:
\begin{equation}
    \label{eq:2}
\, _FT^{\mu  \nu }=\frac{1}{\mu _0}\left(F^{\mu  \beta }F_{\text{   }\beta }^{\nu }-\frac{1}{4}F_{\alpha  \beta }F^{\alpha  \beta }g^{\mu  \nu }\right)
\end{equation}
When using Lorentz gauge $\nabla _{\nu }A^{\nu }=0$, Maxwell equations can be expressed as $\nabla _{\nu }\nabla ^{\nu }A^{\mu }=-\mu _0 j^{\mu } $. For linear insulating media, when there is no net free charge, there is $A^{0 }=0$. Maxwell equations can be further expressed as $n^2\nabla _0\nabla ^0A^i+\nabla _j\nabla ^jA^i=0$, where $n$ is the refractive index of the media. For simplicity, we will only discuss the plane wave solution in the form of $\left(A^{\mu }\right)=\left(0,0,\psi\sin\left(-\frac{\omega  x^0}{c}\pm\frac{\omega n  x^1}{c}\right),0\right)$. Since the general solution can be decomposed into the superposition of plane wave solutions, the results obtained can be extended to the general case. In this case, the electromagnetic field tensor $\left(F^{\mu  \nu }\right)$ from $\left(F^{\mu  \nu }\right)=\left(\nabla ^{\mu }A^{\nu }-\nabla ^{\nu }A^{\mu }\right)$ can be expressed as:
\begin{equation}
    \label{eq:3}
\left(F^{\mu  \nu }\right)=\left( \begin{array}{cccc}  0 & 0 & 1 & 0 \\  0 & 0 & \pm n & 0 \\  -1 & \mp n & 0 & 0 \\  0 & 0 & 0 & 0 \end{array} \right)\frac{E}{c}
\end{equation}
Where $E= E_0 \cos\left(-\frac{\omega  x^0}{c}\pm\frac{\omega  x^1}{c/n}\right)$ is the electric field intensity and $E_0=\omega \psi$ is the amplitude of the electric field intensity. For the electromagnetic wave incident perpendicular to the dielectric interface, the electric field intensity amplitudes of incident wave, reflected wave and transmitted wave are respectively $\,_{\text{in}}E_0$, $\,_{\text{re}}E_0$ and $\, _{\text{tran}} E_0$. From the boundary conditions of the insulating medium we obtain $\, _{\text{re}}E_0=(1-\frac{2 n \mu _0 }{n  \mu _0 +  \mu  })\left( \, _{\text{in}}E _0\right)$, $\, _{\text{tran}} E_0=\frac{2  \mu  }{n  \mu _0 +  \mu }(\, _{\text{in}}E_0)$.  For the wave propagating in the medium, from Eq.\textbf{(}\ref{eq:1}\textbf{)} and Eq.\textbf{(}\ref{eq:3}\textbf{)}, we obtain the energy-momentum tensor of its pure electromagnetic field is:
\begin{equation}
    \label{eq:4}
(\, _{\text{trF}}T^{\mu  \nu })=\left( \begin{array}{cccc}  \frac{n^2+1 }{2} & n & 0 & 0 \\  n & \frac{n^2+1 }{2} & 0 & 0 \\  0 & 0 & \frac{n^2-1 }{2} & 0 \\  0 & 0 & 0 & -\frac{n^2-1 }{2} \end{array} \right)\frac{(\, _{\text{tran}} E )^2}{c^2\mu _0}
\end{equation}
where $\, _{\text{tran}} E =\, _{\text{tran}} E_0\cos\left(-\frac{\omega  x^0}{c}+\frac{\omega  x^1}{c/n}\right)$.We find that it can be written in the following form:
\begin{equation}
    \label{eq:5}
(\, _{\text{trF}}T^{\mu  \nu })=\left( \begin{array}{cccc}  \varrho c^2-\,_Fp & \varrho c v & 0 & 0 \\  \varrho c v &\varrho v^2+\,_Fp & 0 & 0 \\  0 & 0 & \pm \,_Fp & 0 \\  0 & 0 & 0 & \mp \,_Fp \end{array} \right)
\end{equation}
where $\varrho= \gamma ^2\left(\frac{\,_Fp}{c^2}+\,_F\rho \right)$ and:
\begin{equation}
    \label{eq:6}
\begin{array}{ll}\,_F\rho =\left|\frac{n^2-1}{2 c^4  \mu_0}\left( \, _{\text{tran}}E\right) ^2 \right| , \,_Fp=\left|\frac{n^2-1}{2 c^2  \mu_0}\left( \, _{\text{tran}}E\right) ^2 \right| , \\v=\left\{ \begin{array}{ll} c/n  \text{     }\text{     }\text{     }(n\geq1) \\   cn\text{     }\text{     }\text{     }\text{  }(n<1) \end{array} \right., \gamma=\frac{1}{\sqrt{1-\frac{v^2}{c^2}}}\end{array}
\end{equation}
It can be seen that it is very similar to the energy-momentum tensor of ordinary fluid, except that the pressure may be positive or negative, which is related to the polarization of electromagnetic wave. Where density $\,_F\rho$ and pressure $\,_Fp$ are scalars, which do not change with the transformation of the coordinate system, and their value is $0$ in the free space of $n=1$. $v$ is the propagation speed of energy or information, which cannot be greater than $c$. This is understandable because electromagnetic waves can be seen as a fluid composed of photons. The pressure here is a completely different concept from the traditional light pressure. In the free space infinitely close to the interface $x^1=0$ outside the dielectric interface, the electromagnetic field is the superposition of the incident field and the reflected field. The energy-momentum tensor $(\, _{\text{inreF}}T^{\mu  \nu })$ of the pure electromagnetic field can be obtained from Eq.\textbf{(}\ref{eq:2}\textbf{)} and Eq.\textbf{(}\ref{eq:3}\textbf{)}, which can also be written in the form of Eq.\textbf{(}\ref{eq:5}\textbf{)}, where:
\begin{equation}
    \label{eq:7}
\begin{array}{ll}\,_F\rho =\left|\frac{2\text{  }\left(n \mu _0-\mu \right) }{c^4 \mu _0 \left(n \mu _0+\mu \right)}\left(\, _{\text{in}}E_0\cos \frac{\omega  x^0}{c}\right){}^2\right|, \,_Fp=\rho c^2 , \\v=\left\{ \begin{array}{ll} \frac{c \mu }{n \mu _0}\text{     }\text{     }\text{     }\text{     }\text{     }\text{     }\text{}(\frac{n \mu _0}{\mu }\geq 1 ) \\ \frac{c n \mu _0}{\mu }\text{     }\text{     }\text{     }\text{     }\text{     }(\frac{n \mu _0}{\mu }< 1) \end{array} \right. \end{array}
\end{equation}
Substituting $\, _{\text{tran}} E_0=\frac{2  \mu  }{n  \mu _0 +  \mu }(\, _{\text{in}}E_0)$ into Eq.\textbf{(}\ref{eq:6}\textbf{)} and comparing it with Eq.\textbf{(}\ref{eq:7}\textbf{)}, we find that when $\mu=\mu_0$, there is $\, _{\text{trF}}T^{\mu  \nu }=(\, _{\text{inreF}}T^{\mu  \nu })$. For non-ferromagnetic media, there is $\mu\approx\mu_0$, so there is $\, _{\text{trF}}T^{\mu  \nu }\approx(\, _{\text{inreF}}T^{\mu  \nu })$, $\, _{\text{trF}}T^{\mu  \nu }$ is almost all of the energy-momentum tensor $(\, _{\text{inreF}}T^{\mu  \nu })$ input into the media from free space. Therefore, for the relevant experiments of non-ferromagnetic media, it may be sufficient to use Eq.\textbf{(}\ref{eq:4}\textbf{)} and Eq.\textbf{(}\ref{eq:5}\textbf{)} to analyze.\par
Due to the pressure $\,_Fp$, $\, _{\text{trF}}T^{0 0}$ is no longer the energy density of the electromagnetic wave. But $c(\, _{\text{trF}}T^{0 1})$ is still the energy flow density $S^1$, so the energy $dE=S^1S\Delta t=c(\, _{\text{trF}}T^{0 1})Sdl/v$ of the electromagnetic wave with a cross-sectional area of $S$ and a length of $dl$.  So $\, _{\text{trF}}T^{0 1}=\frac{dE}{Sdl}\frac{v}{c}$.  Since $Sdl=d\Omega$ is the volume of this electromagnetic wave, $\frac{dE}{Sdl}$ is the energy density $\,_Fw$ of the purely electromagnetic wave. Therefore, Eq.\textbf{(}\ref{eq:5}\textbf{)} can be written as:
\begin{equation}
    \label{eq:8}
(\, _{\text{trF}}T^{\mu  \nu })=\left( \begin{array}{cccc} \,_Fw-\,_Fp & \frac{\,_Fwv}{c} & 0 & 0 \\  \frac{\,_Fwv}{c} & \frac{\,_Fwv^2}{c^2}+\,_Fp & 0 & 0 \\  0 & 0 & \pm \,_Fp & 0 \\  0 & 0 & 0 & \mp \,_Fp \end{array} \right)
\end{equation}
Substituting Eq.\textbf{(}\ref{eq:6}\textbf{)} into $\,_Fw=\gamma ^2\left(\frac{\,_Fp}{c^2}+\,_F\rho \right)c^2$ yields $\,_Fw=\frac{\mu }{\mu _0}\frac{1}{2}\left(\boldsymbol{E}.\boldsymbol{D}+\boldsymbol{B}. \boldsymbol{H}\right)$. Similarly, $(\, _{\text{trF}}T^{0 i})=\frac{\mu }{\mu _0}(\frac{1}{c} \boldsymbol E\times  \boldsymbol H )$ can also be obtained. Let $(\, _{\text{tr}}T^{\mu  \nu })=\frac{\mu_0}{\mu  }(\, _{\text{trF}}T^{\mu  \nu })$, we obtain:
\begin{equation}
    \label{eq:9}
(\, _{\text{tr}}T^{\mu  \nu })=\left( \begin{array}{cccc}  w- p & \frac{ wv}{c} & 0 & 0 \\  \frac{ wv}{c} & \frac{ wv^2}{c^2}+ p & 0 & 0 \\  0 & 0 & \pm  p & 0 \\  0 & 0 & 0 & \mp  p \end{array} \right)
\end{equation}
where $w=\frac{1}{2}\left(\boldsymbol{E}.\boldsymbol{D}+\boldsymbol{B}. \boldsymbol{H}\right)$, $p=\left|\frac{n^2-1}{2 c^2  \mu }\left( \, _{\text{tran}}E\right) ^2 \right| $. Because $w=\frac{1}{2}\left(\boldsymbol{E}.\boldsymbol{D}+\boldsymbol{B}. \boldsymbol{H}\right)$  is the energy density including the polarization energy and magnetization energy of the medium, Eq.\textbf{(}\ref{eq:9}\textbf{)} is the energy-momentum tensor of the contribution of the polarization energy and magnetization energy of the medium.\par
The pillar behind the Minkowski momentum is quantum mechanics. In quantum mechanics, the momentum corresponding to the momentum operator is the canonical momentum which contains the contribution of the interaction term. Therefore, we hope to find an energy-momentum tensor containing the contribution of the interaction term. Since the interaction is already included, it should be conserved. We find this energy-momentum tensor after performing the study. It is:
\begin{equation}
    \label{eq:10}
\,_{FI}T^{\mu  \nu }=\frac{-2}{\sqrt{-g}}\frac{\delta (\,_{FI}\mathcal{L} \sqrt{-g})}{\delta  g_{\mu  \nu }}
\end{equation}
where $\,_{FI}\mathcal{L}=\frac{\mu_0}{\mu  }(\, _F\mathcal{L}+\frac{1}{2}\, _I\mathcal{L})$. In this case, the total Lagrangian density $\mathcal{L}=\, _F\mathcal{L}+\, _I\mathcal{L}+\, _M\mathcal{L}$ is divided into two parts, $\,_{FI}\mathcal{L}$ is closely related to the electromagnetic field and $\mathcal{L}-\,_{FI}\mathcal{L}$ to the medium. From Eq.\textbf{(}\ref{eq:10}\textbf{)}, we obtain:
\begin{equation}
    \label{eq:11}
\begin{array}{ll}\, _ {FI} T^{\mu  \nu }=\frac{1}{\mu}\left(F^{\mu  \beta }F_{\text{   }\beta }^{\nu }-\frac{1}{4}F_{\alpha  \beta }F^{\alpha  \beta }g^{\mu  \nu }\right)\\
\text{     }\text{     }\text{     }\text{     }\text{     }\text{     }\text{     }\text{     }\text{     }\text{     }\text{     }\text{     }+\frac{\mu_0}{2\mu}(-A^{\mu }j^{\nu }-A^{\nu }j^{\mu }+A_{\beta }j^{\beta }g^{\mu  \nu })\end{array}
\end{equation}
Substituting $\left(A^{\mu }\right)=\left(0,0,\psi\sin\left(-\frac{\omega  x^0}{c}\pm\frac{\omega n  x^1}{c}\right),0\right)$, $j^{\mu }=-\frac{1}{\mu_0}\nabla _{\nu }\nabla ^{\nu }A^{\mu }$ and Eq.\textbf{(}\ref{eq:3}\textbf{)} into  Eq.\textbf{(}\ref{eq:11}\textbf{)}, we obtain:
\begin{equation}
    \label{eq:12}
 (\, _ {FI} T^{\mu  \nu })=\left( \begin{array}{cccc}  w-p-\, _Ip & \frac{w v}{c} & 0 & 0 \\  \frac{w v}{c} & \frac{w v^2}{c^2}+p+\, _Ip & 0 & 0 \\  0 & 0 & \pm p\mp \, _Ip & 0 \\  0 & 0 & 0 & \mp p\pm \, _Ip \end{array} \right)
\end{equation}
Where $w= \frac{n^2}{c^2\mu }(\, _{\text{tran}}E)^2 $, $p=\left|\frac{n^2-1}{2 c^2  \mu }(\, _{\text{tran}}E )^2 \right|$, $\, _Ip=\left|\frac{n^2-1}{2 c^2  \mu }(\, _{\text{tran}}E_0)^2\sin ^2\frac{\omega  \left(-x^0+n x^1\right)}{c} \right| $. It can be verified that there is indeed $\partial _{\mu } (\, _ {FI} T^{\mu  \nu })=0$. The pressure $\, _Ip$ is the contribution of the interaction term; $p+\, _Ip$ is the total pressure. Take an average of $(\, _ {FI} T^{\mu  \nu })$ in one cycle, and we get:
\begin{equation}
    \label{eq:13}
\overline{\, _ {FI} T^{\mu  \nu }}=\left( \begin{array}{cccc}  \frac{1}{2c^2 \mu } & \frac{n}{2c^2 \mu } & 0 & 0 \\  \frac{n}{2 c^2\mu  } & \frac{n^2}{2c^2 \mu } & 0 & 0 \\  0 & 0 & 0 & 0 \\  0 & 0 & 0 & 0 \end{array} \right)(\, _{\text{tran}}E_0)^2
\end{equation}
From Eq.\textbf{(}\ref{eq:7}\textbf{)}, the average energy density of the light entering the interface in the free space outside the interface is $\overline{w}=\frac{n^2 \mu _0}{2c^2 \mu ^2}\left( \, _{\text{tran}}E_0 \right)^2$. Since the energy of photons in free space is $\hbar  \omega$, the average photon number density is $\overline{\, _{\gamma }n}=\frac{n^2 \mu _0}{2c^2 \mu ^2}\frac{1}{\hbar  \omega}\left( \, _{\text{tran}}E_0 \right)^2$. Because the volume of the beam outside the medium is $\Omega$, the volume inside the medium becomes $ \frac{\mu_0}{\mu} \Omega$. Therefore, the average photon number density in the medium is $\overline{\, _{\text{$\gamma $tr}}n}=\frac{n^2}{2c^2 \mu \text{  }} \frac{1}{\hbar  \omega }\left( \, _{\text{tran}}E_0 \right)^2$.\par
For the canonical momentum considering the contribution of interaction, the pressure $p+\, _Ip$ should be included, so the average momentum flow density is $\overline{\, _ {FI} T^{1 1 }}=\overline{\frac{w v^2}{c^2}+p+\, _Ip}$. The time required for light with a cross-sectional area of $S$ and a length of $l$ in the medium to pass through the cross-section is $\Delta t=l/v=nl/c$. Therefore, according to $\, _ {M}P^1=\overline{\, _ {FI} T^{1 1}}S\Delta t$, the canonical momentum of light in volume $\Omega=Sl$ is $\, _ {M}P^1=\frac{n^3}{2c^3 \mu }\left( \, _{\text{tran}}E_0 \right)^2\Omega $. Therefore, the average canonical momentum of photons is $\overline{ \, _ {\gamma M}P^1}=\, _ {M}P^1/(\overline{\, _{\text{$\gamma $tr}}n}\Omega)=\frac{n \hbar  \omega}{c}$. It is Minkowski momentum.\par
For the mechanical momentum without considering the contribution of interaction, the pressure $p+\, _Ip$ should not be included, so $\overline{\, _ {FI} T^{1 1 }}$ is not the mechanical momentum flow density, and $\overline{\, _ {FI} T^{1 1 }-(p+\, _Ip)}=\frac{\overline{w} v^2}{c^2} =\frac{1}{2c^2\mu }(\, _{\text{tran}}E_0)^2$ is the mechanical momentum flow density.  Therefore, according to $\, _ {A}P^1=\frac{\overline{w} v^2}{c^2} S\Delta t$, the mechanical momentum of light in volume $\Omega=Sl$ is $\, _ {A}P^1=\frac{n }{2c^3 \mu }\left( \, _{\text{tran}}E_0 \right)^2\Omega $. Therefore, the average mechanical momentum of photons is $\overline{ \, _ {\gamma A}P^1}=\, _ {A}P^1/(\overline{\, _{\text{$\gamma $tr}}n}\Omega)=\frac{ \hbar  \omega}{nc}$. It is the Abraham momentum.\par
So far, we have derived Minkowski momentum and Abraham momentum simultaneously with the same energy-momentum tensor Eq.\textbf{(}\ref{eq:12}\textbf{)}. It can be seen that these two kinds of momentum are compatible rather than conflicting, and they are both correct and effective. Our opinion is the same as Stephen's, but our argument method is completely different. Our argument can better reveal the connection and difference between the two kinds of momentum.\par
From Eq.\textbf{(}\ref{eq:5}\textbf{)}, we can see that the pressure can be positive or negative. Considering the polarization direction of electromagnetic wave, we obtain:
\begin{equation}
    \label{eq:14}
(\, _FT^{\mu  \nu })=\left( \begin{array}{cccc}  w-p & \frac{w v}{c} & 0 & 0 \\  \frac{w v}{c} & w\frac{v^2}{c^2}+p & 0 & 0 \\  0 & 0 & p^{22} & p^{23}  \\  0 & 0 & p^{32}  & p^{33} \end{array} \right)
\end{equation}
Where $p=\left|\frac{n^2-1}{2 c^2 \mu_0 }\left(\, _{\text{tran}}E\right){}^2\right|$, $p^{22}=-p^{33}=p \cos (2 \theta )$, $p^{23}=p^{32}=-p \sin (2 \theta )$. It can be seen that the pressure on the side of the beam is not isotropic, but related to the polarization of the light. In the direction at an angle $\theta$ to the electric field intensity vector, the magnitude of the pressure is $p \cos (2 \theta )$. The existence of the side pressure of the beam is an important difference between our energy-momentum tensor and Minkowski tensor and Abraham tensor. For Eq.\textbf{(}\ref{eq:12}\textbf{)} considering the contribution of the interaction term, the average value of the lateral pressure will be $0$, but the instantaneous value is:
\begin{equation}
    \label{eq:15}
\, _{\text{side}}p=\frac{ n^2-1 }{2 c^2 \mu _0}\left(\, _{\text{tran}}E_0\right){}^2\cos \frac{2 \omega \left(n x^1-x^0\right)}{c}
\end{equation}
which is not $0$. We hope that some experimental physicist can design experiments to verify the existence of the side pressure of the beam.\par
Since the beam can be regarded as a fluid, we can also discuss its Bernoulli effect. It is assumed that the light beam is incident vertically from a medium with a refractive index of $\, _1n$ to a medium with a refractive index of $\, _2n$. From Eq.\textbf{(}\ref{eq:5}\textbf{)}, we obtain:
\begin{equation}
    \label{eq:16}
\, _{2F}p=\frac{4 \left(\, _1n\right){}^2 }{ \left(\, _1n+\, _2n\right){}^2}\frac{\left(\, _2n^2-1\right)}{\left(\, _1n^2-1\right) }\left(\, _{1F}p\right)
\end{equation}
Therefore, the greater the refractive index of the medium, the greater the pressure, because the greater the refractive index, the smaller the speed of energy propagation. The lower the velocity, the higher the pressure is a characteristic of the Bernoulli effect. However, Eq.\textbf{(}\ref{eq:16}\textbf{)} is not suitable for describing light incident from free space. The pressure of light in free space is $0$. From Eq.\textbf{(}\ref{eq:5}\textbf{)}, we obtain that the pressure of light in medium is:
\begin{equation}
    \label{eq:17}
\, _{ F}p=\frac{2(n-1)}{n+1}\left(\frac{\, _{in}S^1}{c}\right)
\end{equation}
where $ \, _{in}S^1$ is the energy flow density of the incident beam. Eq.\textbf{(}\ref{eq:17}\textbf{)} can be used to explain Ashkin's experiment\textcolor[rgb]{0.184313725,0.188235294117647,0.564705882}{\cite{1973Radiation}}. Although Leonhardt believes that Ashkin's experimental results have other reasons\textcolor[rgb]{0.184313725,0.188235294117647,0.564705882}{\cite{Leonhardt}}. The pressure given in this paper has a significant effect on the results of Ashkin experiment, so the experiment tends to support Minkowski momentum.\par
For Jones' experiment of light pressure in the medium in 1951\textcolor[rgb]{0.184313725,0.188235294117647,0.564705882}{\cite{JONES}}, we use Eq.\textbf{(}\ref{eq:4}\textbf{)} and incident and reflected waves that meet the metal boundary conditions\textcolor[rgb]{0.184313725,0.188235294117647,0.564705882}{\cite{Brevik1979}}, and assume that the deflection angle is proportional to the pressure on the reflector, that is, there is $\overline{\,_FT^{11}}=\kappa \theta$, and we get that the deflection angle is:
\begin{equation}
    \label{eq:18}
\theta\approx \frac{-k+\sqrt{k^2+4 a b \chi ^4}}{2 a \chi ^2}
\end{equation}
where $a\approx4 n^2\left( \sqrt{R}\cos  \delta +2R\right)$, $b\approx1+R-2 \sqrt{R} \cos  \delta +n^2\left(1+R+2 \sqrt{R} \cos  \delta \right)$, $k=\frac{4\kappa c^2 \mu _0}{(\, _{\text{in}}E_0)^2}$, $\chi =\frac{2 (\,_{\text{gla}}n)}{n+ (\,_{\text{gla}}n)}$. Where $\kappa$ is a parameter related to the elasticity of the thin wire used in the experiment, and $\, _{\text{gla}}n$ is the refractive index of the glass used in the experiment. The reflectivity of rhodium metal is about $R\approx0.7$. The phase factor $\delta$ can be estimated by $\delta =\text{arctan}(-0.1)$\textcolor[rgb]{0.184313725,0.188235294117647,0.564705882}{\cite{Brevik1979}}. The power of the light source used in the experiment is known to be $\, _{\text{in}}P=30W$\textcolor[rgb]{0.184313725,0.188235294117647,0.564705882}{\cite{Brevik1979}}. Based on the boundary conditions between air and glass, we obtain $ \, _{\text{in}}E_0 =\frac{2}{ {  \,_{\text{gla}}n }+1}\sqrt{c \mu _0\left( \left.\, _{\text{in}}P\right/S\right)}$, where $S$ is the cross-sectional area of the light beam emitted by the light source. According to the above data, we get the curve of $\, _n\theta/\, _{\text{air}}\theta$ versus refractive index $n$ as shown in (\hyperref[fig:1]{fig.1}):
\begin{figure}[h]
\centering
\includegraphics[scale=0.95]{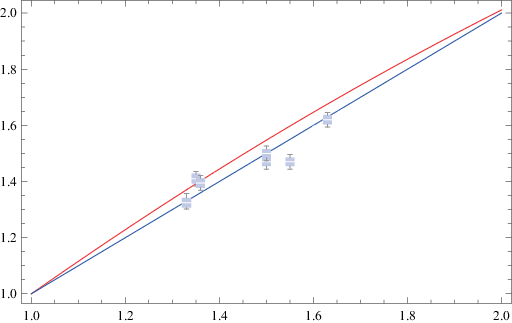}
\caption{Change curve of $\, _n\theta/\, _{\text{air}}\theta$ with $n$. The red line is $\, _n\theta/\, _{\text{air}}\theta$, and the blue line is $n$. Where $\kappa =0.01\sim500Pa/rad$ and $S=5\sim0.0001m^2$. Box-plot is Jones' experimental data.}
\label{fig:1}
\end{figure}\\
The theoretical values are in good agreement with the experimental results. The pressure given in this paper has a significant effect on the results of Jones experiment, so the experiment tends to support Minkowski momentum.\par
As for the optical fiber deformation experiment conducted by Weilong in 2008\textcolor[rgb]{0.184313725,0.188235294117647,0.564705882}{\cite{Weilong2008}}, we applied Eq.\textbf{(}\ref{eq:4}\textbf{)},Eq.\textbf{(}\ref{eq:5}\textbf{)} and boundary conditions to obtain that the force applied on the free end of the optical fiber is:
\begin{equation}
    \label{eq:19}
\overline{ f^1}\approx\frac{2 (n-1)}{ c n(n+1)  } \overline{ \, _{\text{tran}}P }
\end{equation}
Here $\overline{ \, _{\text{tran}}P }$ is the average power of light at the free end of the optical fiber. Our results are consistent with Weilong's, but the energy-momentum tensor we used is not Abraham tensor. The influence of the pressure given in this paper on the results of Weilong experiment can be ignored, so Weilong experiment supports the Abraham momentum.\par
Summarizing, we obtain the following result. When $p+\, _Ip$ in Eq.\textbf{(}\ref{eq:12}\textbf{)} is included in the momentum flow density, Minkowski momentum is obtained. On the contrary, we get the Abraham momentum. These two kinds of momentum are compatible rather than opposite. For general experiments in non-ferromagnetic media, it is sufficient to use Eq.\textbf{(}\ref{eq:4}\textbf{)},Eq.\textbf{(}\ref{eq:5}\textbf{)} for analysis due to $\mu\approx\mu_0$. When the pressure significantly affects the experimental results, the experiment will tend to support Minkowski momentum, and vice versa. The findings of this paper provide a new way to solve the Abraham-Minkowski dilemma.

\bibliographystyle{apsrev4-1}
\bibliography{reference}

\end{document}